\documentclass[reqno,11pt]{amsart}

\usepackage{amssymb,epsf}

\def\be{\begin{equation}}
\def\ee{\end{equation}}
\def\ba{\begin{align}}
\def\bm{\begin{multline}}
\def\bfig{\begin{figure}[htb]}
\def\efig{\end{figure}}

\numberwithin{equation}{section}
\newtheorem{theorem}{Theorem}

\newcommand{\nn}{\nonumber}

\DeclareMathSymbol{\leqslant}{\mathalpha}{AMSa}{"36}
\DeclareMathSymbol{\geqslant}{\mathalpha}{AMSa}{"3E}
\DeclareMathSymbol{\doteqdot}{\mathalpha}{AMSa}{"2B}
\DeclareMathSymbol{\circlearrowright}{\mathalpha}{AMSa}{"08}
\DeclareMathSymbol{\subsetneq}{\mathalpha}{AMSb}{"28}
\DeclareMathSymbol{\supsetneq}{\mathalpha}{AMSb}{"29}
\renewcommand{\leq}{\;\leqslant\;}
\renewcommand{\geq}{\;\geqslant\;}

\newcommand{\dd}{{\rm d}}
\newcommand{\e}[1]{\,{\rm e}^{#1}\,}

\newcommand{\sumtwo}[2]{\sum_{\substack{#1 \\ #2}}}

\def\Tr{{\operatorname{Tr\,}}}

\newcommand{\expval}[1]{\langle #1 \rangle}

\newcommand{\upchi}{\raise 2pt \hbox{$\chi$}}

\def\writefig#1 #2 #3 {\rlap{\kern #1 truecm \raise #2 truecm
\hbox{#3}}}

\newcommand{\caH}{{\mathcal H}}

\newcommand{\caU}{{\mathcal U}}

\newcommand{\caX}{{\mathcal X}}

\newcommand{\bbN}{{\mathbb N}}

\newcommand{\bbR}{{\mathbb R}}

\newcommand{\bbT}{{\mathbb T}}

\newcommand{\bbZ}{{\mathbb Z}}

\newcommand{\bsc}{{\boldsymbol c}}

\newcommand{\bsf}{{\boldsymbol f}}

\newcommand{\bsp}{{\boldsymbol p}}

\newcommand{\bsP}{{\boldsymbol P}}

\newcommand{\bsrho}{{\boldsymbol\rho}}
\newcommand{\bsvarrho}{{\boldsymbol\varrho}}
\newcommand{\bssigma}{{\boldsymbol\sigma}}

\begin{document}


\title{Feynman cycles in the Bose gas}

\author{Daniel Ueltschi}

\address{Daniel Ueltschi \hfill\newline
Department of Mathematics \hfill\newline
University of Arizona \hfill\newline
Tucson, AZ 85721, USA\hfill\newline
}
\email{ueltschi@email.arizona.edu}

\maketitle

\begin{quote}
{\small
{\bf Abstract.} We study the lengths of the cycles formed by trajectories in the
Feynman-Kac representation of the Bose gas. We discuss the occurrence of infinite cycles
and their relation to Bose-Einstein condensation.
}  

\vspace{1mm}
\noindent
{\footnotesize {\it Keywords:} Bose-Einstein condensation. Feynman cycles. Infinite cycles.}

\vspace{1mm}
\noindent
{\footnotesize {\it 2000 Math.\ Subj.\ Class.:} 82B10, 82B21, 82B26, 82D50.\\
{\it PACS numbers:} 03.75.Hh, 05.30.-d, 05.30.Jp, 05.70.Fh, 31.15.Kb}
\end{quote}

\section{Introduction}

Bose and Einstein understood 80 years ago that a curious phase transition occurs in a gas
of non-interacting bosons; it is now commonly refered to as Bose-Einstein condensation. Real particles interact, however, and for many years there
were doubts that this transition takes place in natural systems. London
suggested in 1938 that superfluid Helium undergoes a Bose-Einstein condensation, and this
idea is largely accepted nowadays. Bogolubov considered interacting systems; careful
approximations allowed him to get back to a non-interacting gas, but with a different
dispersion relation. See \cite{ZB} and \cite{LSSY} for more discussion and partial
justifications of Bogolubov theory.

In 1953 Feynman studied the system in the Feynman-Kac representation
\cite{Fey}. The partition function can be expanded as a gas of
trajectories living in $(d+1)$ dimensions. The extra dimension is
commonly refered to as ``the time'', although it is not related to
physical time. The situation is illustrated in Fig.\
\ref{figfeykac}. A finite system with $N$ particles induces a
probability on the group $S_N$ of permutations of $N$ elements.
Feynman considered the probability for a given particle to belong to
a cycle of length $n$. In the thermodynamic limit, there may be
strictly positive probability for infinite cycles to be present, and
Feynman suggested to use this as an order parameter for
Bose-Einstein condensation.

\bfig \epsfxsize=100mm \centerline{\epsffile{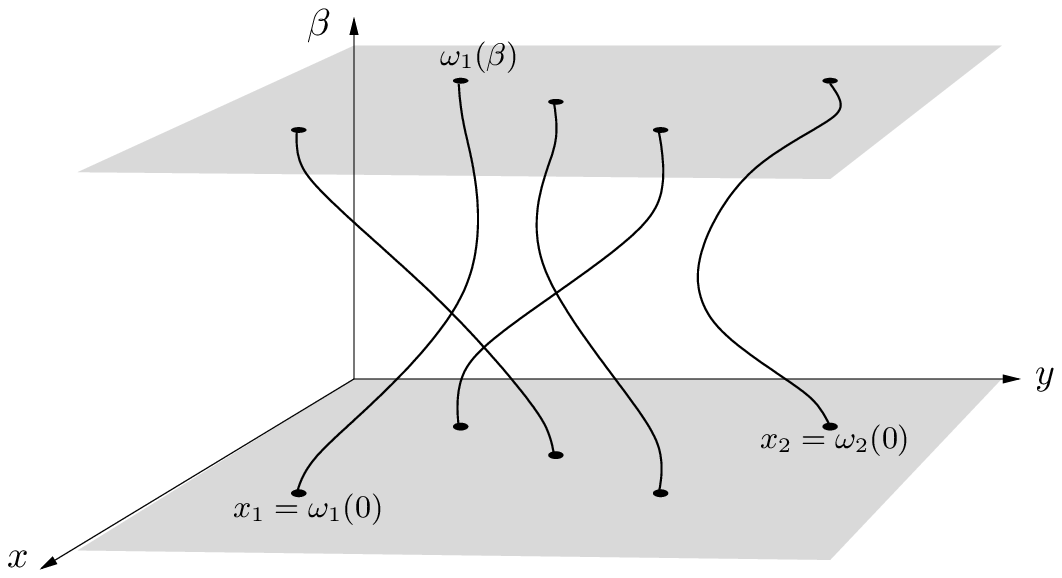}} \caption{The
Feynman-Kac representation of the partition function for a gas of
bosons. The horizontal plane represents the $d$ spatial dimensions,
and the vertical axis is the imaginary time dimension. The picture
shows a situation with five particles and two cycles, of respective
lengths 4 and 1.} \label{figfeykac}
\end{figure}

A few years later, in 1956, Penrose and Onsager introduced the concept of ``off-diagonal
long-range order'' \cite{PO}. Formally, it is a correlation between positions $x$ and $y$
given by $\sigma(x,y) = \langle c^\dagger(x) c(y) \rangle$.
The system displays off-diagonal long-range order when this correlation is strictly
positive, uniformly in the size of the system and in $|x-y|\to\infty$. One can write a Feynman-Kac version of this correlation, and
it involves a special cycle starting at $x$ and ending at $y$; this cycle may wind many times
around the imaginary time direction. In the limit where $x$ and $y$ are infinitely distant there
corresponds a notion of infinite open cycle that is reminiscent of Feynman's approach.

Feynman's order parameter is simpler; it is often used in numerical
simulations or in order to gain heuristic understanding. On the
other hand, everybody agrees that Penrose and Onsager order
parameter is the correct one. Surprisingly, the question of their
equivalence is usually eluded, and many physicists implicitely
assume equivalence to hold. The first mathematical investigation of
this question is due to S\"ut\H o, who showed that equivalence holds
in the ideal gas. Indeed, he proved that infinite cycles occur in
the presence of condensation \cite{Suto}, and that no infinite
cycles occur in the absence of condensation \cite{Suto2}; the latter
result uses probabilistic methods from the theory of large
deviations. These results have been extended to mean-field systems
in \cite{BCMP} and \cite{DMP}.

In this paper we explore the links between Feynman cycles and off-diagonal long-range
order. Let $\bssigma(x)$ denote the off-diagonal correlation between the origin and
$x\in\bbR^d$, and $\bsvarrho(n)$ denote the density of particles in cycles of length $n$. We
propose the following formula that relates both concepts:
\be
\label{LaFormule}
\bssigma(x) = \sum_{n\geq1} \bsc_n(x) \bsvarrho(n) + \bsc_\infty(x) \bsvarrho(\infty).
\ee

Mathematically, the problem is not well posed. Many choices for the coefficients $\bsc_n$
are possible --- a trivial choice is $\bsc_n(x) = \bssigma(x)/\rho$ for all $n$, including
$n=\infty$. We will see, however, that there is a natural definition for $\bsc_n(x)$ in
terms of Wiener trajectories. In any case, we conjecture that Eq.\ \eqref{LaFormule} holds with coefficients
satisfying
$$
0 \leq \bsc_n(x) \leq 1, \quad\quad 0 \leq \bsc_\infty(x) \leq 1,
$$
for all $n,x$. In addition, we should have
$$
\lim_{n\to\infty} \bsc_n(x) = \bsc_\infty(x)
$$
for any $x$, and
$$
\lim_{|x|\to\infty} \bsc_n(x) = 0
$$
for any finite $n$, but not uniformly in $n$; $\bsc_\infty(x)$ may
converge to a strictly positive constant $\bsc$. If $\bsc=1$, we get
from the dominated convergence theorem that $\lim_{|x|\to\infty}
\bssigma(x) = \bsvarrho(\infty)$ --- in which case the off-diagonal
long-range order parameter is equal to the density of cycles of
infinite lengths.

We establish this formula and these properties in the case of the ideal gas, where we show that
\be
\bsc_n(x) = \e{-x^2/4n\beta}, \quad\quad \bsc_\infty(x) = 1.
\ee
We discuss the validity of the formula \eqref{LaFormule} in the interacting gas, proving
that these properties hold true in a regime without Bose-Einstein condensation. The two
order parameters should not be always equivalent, however. It is argued in \cite{Uel2} that
they differ when the bosons undergo a regular condensation into a crystalline phase. There
is no off-diagonal long-range order, but infinite cycles may be present.

We work in the Feynman-Kac representation of the Bose gas. This representation is
standard, see e.g.\ \cite{Far} for a clear and concise review, and \cite{Gin} for a
complete introduction. We assume the reader to possess
some familiarity with it and in Section \ref{seccyc} we directly define the main
expressions --- partition functions, density of cycles, off-diagonal long-range order --- in
terms of space-time trajectories. But basic notions and properties are reviewed in Appendix
\ref{secFKrep}.

The situation simplifies in absence of interactions; we consider the
ideal gas in Section \ref{secidealgas}, where we state and prove the
formula that relates the two order parameters. The ideal gas is best
discussed in the canonical ensemble. Rigorous proofs of macroscopic
occupation of the zero mode have been proposed and they involve the
grand-canonical ensemble, with a chemical potential that depends on
the volume. Appendix \ref{secgazparfait} proposes a simple proof in
the canonical ensemble.

Interacting systems constitute a formidable challenge; they are discussed in
Section \ref{secintgas}, where partial results are obtained.

In this paper, we denote {\it finite volume} expressions in {\it plain characters}, and
{\it infinite volume} expressions in {\it bold characters}. Further, we always consider
the canonical and grand-canonical ensembles where the temperature $1/\beta$ is fixed; we
alleviate the notation by omitting the $\beta$ dependence of all quantities.

\section{Feynman cycles and off-diagonal long-range order}
\label{seccyc}

\subsection{Partition functions}

Our Bose gas occupies a $d$-dimensional domain $D$, always a cubic box of size $L$ and
volume $V=L^d$. We consider periodic boundary conditions. Let $\rho$ denote the particle
density, $\beta$ the inverse temperature, and $\mu$ the chemical potential. The canonical partition function in the
Feynman-Kac representation is given by
\bm
\label{FKrepcanpartfct}
Y(N) = \sum_{k=1}^N \frac1{k!} \sumtwo{n_1,\dots,n_k \geq 1}{n_1 + \dots + n_k = N} \int_{D^k} \dd x_1 \dots \dd x_k \int \dd
W^{n_1 \beta}_{x_1 x_1}(\omega_1) \dots \dd W^{n_k \beta}_{x_k x_k}(\omega_k) \\
\biggl[ \prod_{j=1}^k \frac1{n_j} \e{-\beta \caU(\omega_j)} \biggr]
\prod_{1\leq i<j\leq k} \e{-\beta \caU(\omega_i,\omega_j)}.
\end{multline}
This expression is illustrated in Fig.\ \ref{figfeykac}. In words, we sum over the number
$k$ of closed trajectories and over their respective winding numbers $n_1,\dots,n_k$. We
integrate over the initial positions $x_1,\dots,x_k$. We integrate over trajectories
$\omega_j : [0,n_j \beta] \to D$ that start and end at $x_j$; here, $W^\beta_{xx}$ denotes the Wiener
measure. See Appendix \ref{secFKrep} for more information, and in particular Eq.\
\eqref{Wienerperiodic} for the normalization condition. Trajectories wind around the
time direction according to their winding numbers; because of periodic boundary
conditions, they may also wind around space directions.

Given a trajectory $\omega$ with winding number $n$, the function
$\caU(\omega)$ denotes the interactions between different legs;
explicitly,
\be
\caU(\omega) = \sum_{0\leq i<j\leq n-1} \frac1\beta
\int_0^\beta U \bigl( \omega(i\beta+s) - \omega(j\beta+s) \bigr) \dd
s.
\ee
And $\caU(\omega,\omega')$ denotes the interactions between
closed trajectories $\omega$ and $\omega'$, of respective winding
numbers $n$ and $n'$:
\be
\caU(\omega,\omega') = \sumtwo{0\leq i\leq
n-1}{0\leq j\leq n'-1} \frac1\beta \int_0^\beta U \bigl(
\omega(i\beta+s) - \omega'(j\beta+s) \bigr) \dd s.
\ee
The function $U(x)$ represents the pair interaction potential
between two particles separated by a distance $|x|$. We suppose that
$U(x)$ is nonnegative and spherically symmetric. We can allow the
value $+\infty$; all that is needed is that $\e{-\beta
\caU(\omega)}$ and $\e{-\beta \caU(\omega,\omega')}$ be measurable
functions with respect to the Wiener measure --- any piecewise
continuous function $D \to [0,\infty]$ can be considered at this
point.

The grand-canonical partition function is \be
\label{FKrepgdcanpartfct} Z(\mu) = \sum_{N\geq0} \e{\beta\mu N} Y(N)
\ee (with the understanding that $Y(0)=1$). We also need partition
functions where a given trajectory $\omega_0$ is present --- these
will be needed in the expression for cycle densities, see
\eqref{densitycycles} and \eqref{densitycyclesgdcan}. Namely, we
define \bm \label{fpartaveccycle} Y(N;\omega_0) = \sum_{k=1}^N
\frac1{k!} \sumtwo{n_1,\dots,n_k \geq 1}{n_1+\dots+n_k=N} \int_{D^k}
\dd x_1 \dots \dd x_k \int \dd W_{x_1 x_1}^{n_1 \beta}(\omega_1)
\dots \dd W_{x_k x_k}^{n_k
\beta}(\omega_k) \\
\biggl[ \prod_{j=1}^k \frac1{n_j} \e{-\beta \caU(\omega_j)} \biggr]
\prod_{0\leq i<j\leq k} \e{-\beta \caU(\omega_i,\omega_j)}.
\end{multline}
The dependence on $\omega_0$ comes from the last term, where the
product includes terms with $i=0$. Notice that
\be
Y(N;\omega) \leq Y(N),
\ee
with equality iff $U(x)\equiv0$, i.e.\ in absence of interactions. Finally, we introduce
\be
\label{fpartgdcanaveccycle}
Z(\mu,\omega) = \sum_{N\geq0} \e{\beta\mu N} Y(N,\omega)
\ee
(we set $Y(0,\omega_0)=1$).
We also have $Z(\mu,\omega) \leq Z(\mu)$, with equality iff $U(x)\equiv0$.

\subsection{Cycle lengths}

We now introduce the density of particles in cycles of length $n$, both in the canonical
and grand-canonical ensembles. We denote the particle density by $\rho = \frac NV$. When
discussing the canonical ensemble, we
always suppose that $\rho$ and $V$ are such that $N=\rho V$ is an integer. The number of particles in cycles of length $n$ is given by
the random variable $\sum_{j=1}^k n \delta_{n_j,n}$.
Averaging over all configurations of space-time closed trajectories, we get
\ba
\varrho_\rho(n) = &\frac1{Y(N)} \sum_{k=1}^N \frac1{k!} \sumtwo{n_1,\dots,n_k \geq
1}{n_1+\dots+n_k=N} \biggl[ \frac1V \sum_{j=1}^k n \delta_{n_j,n} \biggr] \int_{D^k} \dd
x_1 \dots \dd x_k \nn\\
&\hspace{15mm} \int \dd W^{n_1 \beta}_{x_1 x_1}(\omega_1) \dots \dd
W^{n_k \beta}_{x_k x_k}(\omega_k) \biggl[ \prod_{j=1}^k \frac1{n_j}
\e{-\beta \caU(\omega_j)} \biggr] \prod_{1\leq i<j\leq k}
\e{-\beta \caU(\omega_i,\omega_j)} \nn\\
= &\int \dd W_{00}^{n\beta}(\omega) \e{-\beta \caU(\omega)}
\frac{Y(N-n;\omega)}{Y(N)}. \label{densitycycles}
\end{align}
The last line follows from the first line by replacing $\sum_{j=1}^k
n \delta_{n_j,n}$ with $nk \delta_{n_1,n}$; isolating the integral
over $\omega_1$; using the definition \eqref{fpartaveccycle}; using
translation invariance, and $\int_D \dd x_1 = V$. Similarly, we have
the grand-canonical expression
\be
\label{densitycyclesgdcan}
\varrho_\mu(n) = \e{\beta\mu n} \int\dd W_{00}^{n\beta}(\omega)
\e{-\beta \caU(\omega)} \frac{Z(\mu;\omega)}{Z(\mu)}.
\ee

One easily checks that
\[
\begin{split}
&\sum_{n\geq1} \varrho_\rho(n) = \tfrac NV \equiv \rho, \\
&\sum_{n\geq1} \varrho_\mu(n) = \langle \tfrac NV \rangle \equiv \rho(\mu).
\end{split}
\]
We consider the thermodynamic limits of $\varrho_\rho(n)$ and $\varrho_\mu(n)$. Since $0 \leq
\varrho_\rho(n) \leq \rho$, and since $n$ is a discrete index, Cantor diagonal process yields
the existence of a sequence of increasing volumes $V_k$, with $\rho V_k = N_k$ an
integer, such that $\varrho_\rho(n)$ converges to some limit that we denote
$\bsvarrho_\rho(n)$. Similarly, we also obtain the infinite volume limit
$\bsvarrho_\mu(n)$. Fatou's lemma implies that
\be
\sum_{n\geq1} \bsvarrho_\rho(n) \leq \rho, \quad\quad \sum_{n\geq1} \bsvarrho_\mu(n) \leq
\bsrho(\mu).
\ee
This suggests to define the density of particles in infinite cycles by
\be
\label{defrhoinfinity}
\begin{split}
&\bsvarrho_\rho(\infty) = \rho - \sum_{n\geq1} \bsvarrho_\rho(n), \\
&\bsvarrho_\mu(\infty) = \bsrho(\mu) - \sum_{n\geq1} \bsvarrho_\mu(n).
\end{split}
\ee
The main question is whether $\bsvarrho(\infty)$ differs from zero at given
temperature, and at given density or chemical potential.

We chose to discuss densities of particles in cycles of given
length, but one may consider probabilities as well. Namely, we could
introduce the probability for particle 1 to belong to a cycle of
length $n$; it is given by
\be
P_\rho(n) = \int_D \dd x \int\dd
W_{xx}^{n\beta}(\omega) \e{-\beta \caU(\omega)}
\frac{Y(N-n;\omega)}{N \, Y(N)}.
\ee
Thus $\varrho_\rho(n) = \rho
P_\rho(n)$ in the canonical ensemble, and $\bsvarrho_\rho(\infty) =
\rho \bsP_\rho(\infty)$. Things are not so simple in the
grand-canonical ensemble. The probability $P_{\mu}(n)$ is
\be
\label{invonvenientprob}
P_{\mu}(n) = \int_D \dd x \int \dd
W_{xx}^{n\beta}(\omega)  \frac{\e{\beta\mu n}}n \e{-\beta
\caU(\omega)} \frac{Z'(\mu;\omega)}{Z(\mu)}.
\ee
Here,
$Z'(\mu;\omega)$ is like $Z(\mu;\omega)$ given in Eqs
\eqref{fpartgdcanaveccycle} and \eqref{fpartaveccycle}, but with a
factor $\frac1{(k+1)!}$ instead of $\frac1{k!}$. Heuristically, we
should have $\langle \frac V{nk} \rangle = 1/\rho(\mu)$, and
$\varrho_\mu(n) = \rho(\mu) P_{\mu}(n)$, but this does not seem easy
to establish. The ratio of partition functions in
\eqref{invonvenientprob} is more difficult to control than the one
in \eqref{densitycyclesgdcan}. We therefore abandon probabilities
and discuss densities.

\subsection{Off-diagonal long-range order}

Let us turn to Penrose and Onsager off-diagonal long-range order. Its Feynman-Kac
representation involves an open trajectory that starts at $x$ and ends at $y$, that
possibly winds several times around the time direction. Precisely, we introduce
\be
\label{odlro}
\begin{split}
&\sigma_\rho(x) = \sum_{n=1}^N \int \dd W_{0x}^{n\beta}(\omega)
\e{-\beta \caU(\omega)}
\frac{Y(N-n;\omega)}{Y(N)}; \\
&\sigma_\mu(x) = \sum_{n\geq1} \e{\beta\mu n} \int \dd
W_{0x}^{n\beta}(\omega) \e{-\beta \caU(\omega)}
\frac{Z(\mu;\omega)}{Z(\mu)}.
\end{split}
\ee
Thermodynamic limits are denoted $\bssigma_\rho(x)$ and $\bssigma_\mu(x)$, provided
they exist! One may actually restrict $\sigma_\rho(x)$ and $\sigma_\mu(x)$ on rational
$x$, and use the Cantor diagonal process to get convergence on a subsequence of
increasing volumes. This is not necessary in this paper, as the limits will be shown to
exist in the regimes of parameters under consideration.

Similarities between Eqs \eqref{densitycycles}, \eqref{densitycyclesgdcan} on the one hand, and
Eqs \eqref{odlro} on the other hand, are manifest. We can write
\be
\label{coefficients}
\begin{split}
&\sigma_\rho(x) = \sum_{n=1}^N c_{n,\rho}(x) \varrho_\rho(n), \\
&\sigma_\mu(x) = \sum_{n\geq1} c_{n,\mu}(x) \varrho_\mu(n),
\end{split}
\ee
where the coefficients $c_{n,\rho}$, $c_{n,\mu}$ are given by
\be
\label{defcoeff}
\begin{split}
&c_{n,\rho}(x) = \biggl[ \int\dd W_{00}^{n\beta}(\omega) \e{-\beta
\caU(\omega)} \frac{Y(N-n;\omega)}{Y(N)} \biggr]^{-1} \int\dd
W_{0x}^{n\beta}(\omega) \e{-\beta \caU(\omega)}
\frac{Y(N-n;\omega)}{Y(N)}, \\
&c_{n,\mu}(x) = \biggl[ \int\dd W_{00}^{n\beta}(\omega) \e{-\beta
\caU(\omega)} \frac{Z(\mu;\omega)}{Z(\mu)} \biggr]^{-1} \int\dd
W_{0x}^{n\beta}(\omega) \e{-\beta \caU(\omega)}
\frac{Z(\mu;\omega)}{Z(\mu)}.
\end{split}
\ee
As above, we denote the thermodynamic limits by $\bsc_{n,\rho}(x)$ and $\bsc_{n,\mu}(x)$,
provided they exist. One should be careful when sending the volume to infinity in Eqs
\eqref{coefficients}, because a ``leak to infinity'' may yield a term involving
$\bsvarrho(\infty)$ --- this actually occurs in the ideal gas, as shown in the next section.

\section{The ideal gas}
\label{secidealgas}

The ideal gas of quantum bosons is fascinating. Particles do not interact, yet they
manage to display a phase transition. Historically, the Bose-Einstein condensation is the
first theoretical description of a phase transition. The ideal gas has been the object of
many studies over the years; let us mention \cite{ZUK,LLS,PZ}. A simple proof of
macroscopic occupation of the zero Fourier mode is presented in Appendix
\ref{secgazparfait}.

In this section we elucidate the relation between cycle lengths and off-diagonal long-range order, thus
clarifying results that were previously obtained by S\"ut\H o \cite{Suto,Suto2}. We work
in the canonical ensemble and establish the formula \eqref{LaFormule} explicitely, for any
dimension $d\geq1$.

\begin{theorem}
\label{thmidealrelation}
For any $0<\beta,\rho<\infty$, there exists a sequence of increasing cubes for which
the thermodynamic limits of $\sigma_\rho(x)$, $c_{n,\rho}(x)$, $\varrho_\rho(n)$ exist for all
$x \in \bbR^d$ and $n \in \bbN$. Further, we have
$$
\bssigma_\rho(x) = \sum_{n\geq1} \e{-\frac{x^2}{4n\beta}} \bsvarrho_\rho(n) +
\bsvarrho_\rho(\infty).
$$
\end{theorem}

The rest of this section is devoted to the proof of Theorem
\ref{thmidealrelation}. The coefficient $c_{n,\rho}(x)$, defined in
Eq.\ \eqref{defcoeff}, has a simpler expression in absence of
interactions. Indeed, we have $\caU(\omega)=0$ and
$Y(N;\omega)=Y(N)$. It follows from properties of the Wiener measure
in periodic boxes, see Eq.\ \eqref{Wienerperiodic}, that
\be
\label{idealcn}
c_{n,\rho}(x) = \sum_{z\in\bbZ^d}
\e{-\frac{L^2}{4n\beta} (\frac xL - z)^2} \bigg/ \sum_{z\in\bbZ^d}
\e{-\frac{L^2}{4n\beta} z^2}.
\ee
Notice that $\lim_{L\to\infty}
c_{n,\rho}(x) = \e{-\frac{x^2}{4n\beta}}$, but the limit is not
uniform in $n$. If the sum over $n$ is restricted to $n \leq cL^2$,
with $c$ any finite constant, we can use the dominated convergence
theorem and we get
\be
\label{petitstermes}
\lim_{L\to\infty}
\sum_{n=1}^{cL^2} c_{n,\rho}(x) \varrho_\rho(n) = \sum_{n\geq1}
\e{-\frac{x^2}{4n\beta}} \bsvarrho_\rho(n).
\ee
(The limit is taken
along the subsequence of increasing volumes for which
$\varrho_\rho(n)$ is known to converge for any $n$.)

We now consider the terms with $cL^2 < n \leq N$. We estimate the sums in \eqref{idealcn} using integrals;
we have
\be
\label{calculusinequality}
\int_{-\infty}^\infty \e{-a(s-b)^2} \dd s - 1 \leq \sum_{k\in\bbZ} \e{-a(k-b)^2} \leq
\int_{-\infty}^\infty \e{-a(s-b)^2} \dd s + 1.
\ee
The Gaussian integral is equal to $\sqrt{\pi/a}$. Consequently,
\be
\biggl[ \frac{\sqrt{4\pi n\beta} - L}{\sqrt{4\pi n\beta} + L} \biggr]^d \leq
c_{n,\rho}(x) \leq \biggl[ \frac{\sqrt{4\pi n\beta} + L}{\sqrt{4\pi n\beta} - L} \biggr]^d.
\ee
These bounds hold provided $\sqrt{4\pi n\beta} > L$. Since $\frac n{L^2} > c$, we have
\be
\biggl[ \frac{\sqrt{4\pi c\beta} - 1}{\sqrt{4\pi c\beta} + 1} \biggr]^d \sum_{n=cL^2}^N
\varrho_\rho(n) \leq \sum_{n=cL^2}^N c_{n,\rho}(x) \varrho_\rho(n)
\leq \biggl[ \frac{\sqrt{4\pi c\beta} +
1}{\sqrt{4\pi c\beta} - 1} \biggr]^d \sum_{n=cL^2}^N \varrho_\rho(n).
\ee
We obtain
\be
\label{grandstermes}
\sum_{n=cL^2}^N c_{n,\rho}(x) \varrho_\rho(n) \leq \biggl[ \frac{\sqrt{4\pi c\beta} +
1}{\sqrt{4\pi c\beta} - 1} \biggr]^d  \biggl( \rho - \sum_{n=1}^{cL^2} \varrho_\rho(n)
\biggr).
\ee
Using \eqref{petitstermes} with $x=0$ and the definition \eqref{defrhoinfinity} of the
density of infinite cycles, we see that the last term converges to
$\bsvarrho_\rho(\infty)$ as $L\to\infty$. It then follows from \eqref{petitstermes} and
\eqref{grandstermes} that
\be
\limsup_{L\to\infty} \sigma_\rho(x) \leq \sum_{n\geq1} \e{-\frac{x^2}{4n\beta}}
\bsvarrho_\rho(n) + \biggl[ \frac{\sqrt{4\pi c\beta} +
1}{\sqrt{4\pi c\beta} - 1} \biggr]^d \bsvarrho_\rho(\infty).
\ee
This inequality holds for any $c$, and the fraction is arbitrarily close to 1 by taking
$c$ large. A lower bound can be derived in a similar fashion, and we obtain the formula stated in Theorem \ref{thmidealrelation}.

\section{The interacting gas}
\label{secintgas}

The interacting gas is much more difficult to study. We prove in this section the absence
of infinite cycles when the chemical potential is negative (Theorem
\ref{thmnoinfinitecycles}). We then study the coefficients
$c_{n,\mu}(x)$ at low density and high temperature, using cluster expansion techniques.
Their thermodynamic limit can be established, and we show that $\bsc_{n,\mu}(x) \to 0$ as
$|x|\to\infty$ (Theorem \ref{thmintgas}).

\begin{theorem}
\label{thmnoinfinitecycles}
Let $0<\beta<\infty$ and $\mu<0$; then
$$
\bsvarrho_\mu(\infty) = 0,
$$
and
$$
\lim_{|x|\to\infty} \limsup_{L\to\infty} \sigma_\mu(x) = 0.
$$
\end{theorem}

\begin{proof}
Since $\caU(\omega)\geq0$ and $Z(\mu;\omega) \leq Z(\mu)$, the
finite volume density $\varrho_\mu(n)$, Eq.\
\eqref{densitycyclesgdcan}, is less than
\be
\varrho_\mu(n) \leq
\e{\beta\mu n} \int \dd W_{00}^{n\beta}(\omega) = \frac{\e{\beta\mu
n}}{(4\pi n\beta)^{d/2}} \sum_{z\in\bbZ^d} \e{-\frac{L^2
z^2}{4n\beta}}.
\ee
The right side is smaller than $\e{\beta\mu n}$
for all $L$ large enough. We can therefore apply the dominated
convergence theorem and we obtain
\be
\rho = \lim_{L\to\infty}
\sum_{n\geq1} \varrho_\mu(n) = \sum_{n\geq1} \bsvarrho_\mu(n).
\ee
It follows that $\bsvarrho_\rho(\infty) = 0$. The statement about
absence of off-diagonal long-range order can be treated similarly.
We have the upper bound
\be
\sigma_\mu(x) \leq \sum_{n\geq1}
\frac{\e{\beta\mu n}}{(4\pi n\beta)^{d/2}} \sum_{z\in\bbZ^d}
\e{-\frac{(x-Lz)^2}{4n\beta}}.
\ee
By dominated convergence,
\be
\limsup_{L\to\infty} \sigma_\mu(x) \leq \sum_{n\geq1}
\frac{\e{\beta\mu n}}{(4\pi n\beta)^{d/2}} \, \e{-x^2 /4n\beta}.
\ee
We can again use the dominated convergence theorem for the limit
$|x|\to\infty$, and we get the claim.
\end{proof}

We continue the study of the interacting gas in the regime where
cluster expansion converges. We assume that the chemical potential
is negative, that the interaction potential $U(x)$ is integrable,
and that the temperature is high enough. The condition in Theorem
\ref{thmintgas} is stronger than necessary, but it is very explicit.
We will invoke a weaker condition in the proof of the theorem that
is based on the ``Koteck\'y-Preiss criterion'' for the convergence
of cluster expansion. Notice that Ginibre's survey \cite{Gin} uses
Kirkwood-Salzburg equations; it applies to a broader range of
potentials, but things are terribly intricate.

\begin{theorem}
\label{thmintgas}
Assume that $\beta$, $\mu$, and $U$ satisfy
$$
\frac1{(4\pi\beta)^{d/2}} \int_{\bbR^d} U(x) \dd x \sum_{n\geq1} n^{-d/2} \leq -\mu.
$$
The thermodynamic limits of $c_{n,\mu}(x)$ and $\varrho_\mu(n)$ exist, and we have
$$
\lim_{|x|\to\infty} \bsc_{n,\mu}(x) = 0
$$
for any $n$.
\end{theorem}

\begin{proof}
We need some notation in order to cast the grand-canonical partition
function in a form suitable for the cluster expansion. Let us
introduce a measure for trajectories that wind arbitrarily many
times around the time direction. Namely, let $\caX_n$ denote the
measure space of continuous trajectories $\omega : [0,n\beta] \to
D$, and let $\caX = \cup_{n\geq1} \caX_n$ be the set of trajectories
in $D$ with arbitrary winding numbers. We introduce the measure
$\nu$ on $\caX$ whose integral means the following:
\be
\int
F(\omega) \dd\nu(\omega) = \sum_{n\geq1} \frac{\e{\beta\mu n}}n
\int_D \dd x \int\dd W_{xx}^{n\beta}(\omega) \e{-\beta \caU(\omega)}
F(\omega).
\ee

It is clear that $\nu$ is a genuine measure on a reasonable measure
space. But we describe the measure $\nu$ with more details for
readers who are interested in analytic technicalities. The
$\sigma$-algebra on $\caX_n$ is the smallest $\sigma$-algebra that
contains the sets $\{ \omega \in \caX_n : \omega(t) \in B \}$, for
any $0\leq t\leq n\beta$, and any Borel set $B \subset D$.
Trajectories of $\caX_1$ can be dilated in the time direction so as
to yield trajectories with arbitrary winding numbers. One can then
consider the product space $\caX_1 \times \bbN$ with the product
$\sigma$-algebra (the $\sigma$-algebra on $\bbN$ being the power
set). The measure of a set of the kind $A \times \{n\}$, with $A$ a
measurable subset of $\caX_1$, is defined as
\be
\nu(A \times \{n\})
= \frac{\e{\beta\mu n}}n \int \dd x \int_{A'} \dd
W_{xx}^{n\beta}(\omega') \e{-\beta \caU(\omega')}.
\ee
Here, we
introduced \be A' = \bigl\{ \omega' \in \caX_n : \omega'(t) =
\omega(nt) \text{ for some } \omega \in A \bigr\}. \ee There is a
unique extension to a measure on $\caX_1 \times \bbN$. There is a
natural correspondence between $\caX$ and $\caX_1 \times \bbN$, and
we consider $\nu$ to be a measure on $\caX$.

With this notation, the grand-canonical partition function
\eqref{FKrepgdcanpartfct} is given by
\be
\label{fpartclexp}
Z(\mu)
= \sum_{k\geq0} \frac1{k!} \int_{\caX^k} \dd\nu(\omega_1) \dots
\dd\nu(\omega_k) \prod_{1\leq i<j\leq k} \Bigl[ \e{-\beta
\caU(\omega_i,\omega_j)} - 1 \Bigr].
\ee
The term $k=0$ is equal to
1 by definition. Then $Z(\mu)$ has exactly the form assumed e.g.\ in
\cite{Uel}. The Koteck\'y-Preiss criterion for the convergence of
the cluster expansion requires the existence of a function $a : \caX
\to \bbR_+$ such that the following inequality holds for any $\omega
\in \caX$:
\be
\label{KPcrit}
\int_\caX \Bigl[ 1 - \e{-\beta
\caU(\omega,\omega')} \Bigr] \e{a(\omega')} \dd\nu(\omega') \leq
a(\omega).
\ee
Choosing $a(\omega) = -\beta\mu n$ (with $n$ the
winding number of the trajectory $\omega$), it was shown in
\cite{Uel} that \eqref{KPcrit} is a consequence of the condition in
Theorem \ref{thmintgas}.

The main result of the cluster expansion is that the partition
function \eqref{fpartclexp} is given by the exponential of a
convergent series. Namely,
\be
\label{fpartexp}
Z(\mu) = \exp
\biggl\{ \sum_{k\geq1} \int_{\caX^k} \dd\nu(\omega_1) \dots
\dd\nu(\omega_k) \, \varphi(\omega_1,\dots,\omega_k) \biggr\}.
\ee
The combinatorial function $\varphi(\omega_1,\dots,\omega_k)$ is
equal to 1 if $k=1$, and is otherwise equal to
\be
\varphi(\omega_1,\dots,\omega_k) = \frac1{k!} \sum_G \prod_{(i,j)
\in G} \Bigl[ \e{-\beta \caU(\omega_i,\omega_j)} - 1 \Bigr].
\ee
The
sum is over {\it connected} graphs with $k$ vertices, and the
product is over edges of $G$. A proof for the relation
\eqref{fpartexp} that directly applies here can be found in
\cite{Uel}.

Observe now that the partition function $Z(\mu;\omega)$ is given by
an expression similar to \eqref{fpartclexp}, where each
$\dd\nu(\omega_j)$ is replaced by $\e{-\beta \caU(\omega,\omega_j)}
\dd\nu(\omega_j)$. Since $\caU(\omega,\omega_j)$ is positive, the
criterion \eqref{KPcrit} is satisfied with this new measure. It
follows that $Z(\mu;\omega)$ has an expansion similar to
\eqref{fpartexp}, and we obtain the following expression for the
ratio of partition functions, \be \label{rapportfpart}
\frac{Z(\mu;\omega)}{Z(\mu)} = \exp \biggl\{ -\sum_{k\geq1}
\int_{\caX^k} \dd\nu(\omega_1) \dots \dd\nu(\omega_k) \Bigl[ 1 -
\prod_{j=1}^k \e{-\beta \caU(\omega,\omega_j)} \Bigr]
\varphi(\omega_1,\dots,\omega_k) \biggr\}. \ee It is not hard to
check that \be
\begin{split}
1 - \prod_{j=1}^k \e{-\beta \caU(\omega,\omega_j)} &= \sum_{j=1}^k
\bigl( 1 - \e{-\beta \caU(\omega,\omega_j)} \bigr) \prod_{i=1}^{j-1}
\e{-\beta
\caU(\omega,\omega_i)} \\
&\leq \sum_{j=1}^k \bigl( 1 - \e{-\beta \caU(\omega,\omega_j)}
\bigr).
\end{split}
\ee Then Equation (5) in \cite{Uel} gives the necessary estimate for
the exponent in \eqref{rapportfpart}, namely \be \sum_{k\geq1}
\int_{\caX^k} \dd\nu(\omega_1) \dots \dd\nu(\omega_k) \Bigl[
\sum_{j=1}^k \bigl( 1 - \e{-\beta \caU(\omega,\omega_j)} \bigr)
\Bigr] \bigl| \varphi(\omega_1,\dots,\omega_k) \bigr| \leq -\beta\mu
n. \ee This bound is uniform in the size of the domain, which is
important. It follows that, as $L\to\infty$, the ratio
$Z(\mu;\omega)/Z(\mu)$ converges pointwise in $\mu$ and $\omega$.
The thermodynamic limits of $c_{n,\mu}(x)$ and $\varrho_\mu(n)$ then
clearly exist. Further, $c_{n,\mu}(x)$ is bounded by \be
c_{n,\mu}(x) \leq \biggl[ \int\dd W_{00}^{n\beta}(\omega) \e{-\beta
\caU(\omega)} \biggr]^{-1} \e{-2\beta\mu n} \int\dd
W_{0x}^{n\beta}(\omega). \ee It is not hard to show that the bracket
is bounded away from zero uniformly in $L$ (but not uniformly in
$\beta$ and $n$). From \eqref{Wienerperiodic}, we have
$$
\lim_{|x|\to\infty} \lim_{L\to\infty} \int\dd W_{0x}^{n\beta}(\omega) = 0.
$$
This implies that $\bsc_{n,\mu}(x)$ vanishes in the limit of infinite $|x|$.
\end{proof}

\section{Conclusion}

We introduced the formula \eqref{LaFormule} that relates the off-diagonal correlation
function and the densities of cycles of given length. This formula involves coefficients
$\bsc_n$ that have a natural definition in terms of integrals of Wiener trajectories. We
conjectured several properties for the coefficients --- these properties can actually be
proved in the ideal gas for all temperatures, and in the interacting gas for high
temperatures. These results seem to indicate that the order parameters of Feynman and
Penrose-Onsager agree. However, heuristic considerations based on the present framework
\cite{Uel2} suggest that, if the gas is in a crystalline phase, the coefficients satisfy
$\bsc_n(x) \leq \e{-a|x|}$ for some $a>0$, and for all $n$ (including $n=\infty$).
Besides, one expects that $\bsvarrho(\infty)>0$ if the temperature is sufficiently low.
The order parameters are not equivalent in this case.

An open problem is to establish the equivalence of the order parameters in weakly
interacting gases in presence of Bose-Einstein condensation. Another question is whether
$\bsc_\infty(x)$ converges, as $|x|\to\infty$, to a number that is strictly between 0 and 1.
The corresponding phase would display a Bose condensate whose density is less than the
density of particles in infinite cycles.

\bigskip
{\bf Acknowledgments:} It is a pleasure to thank Valentin Zagrebnov
and Joe Pul\'e for encouragements, and for pointing out an error
with thermodynamic potentials in the original vresion of the
manuscript. This research was supported in part by the grant
DMS-0601075 from the US National Science Foundation.

\appendix

\section{Feynman-Kac representation of the Bose gas}
\label{secFKrep}

In this appendix we recall some properties of the Wiener measure, and we review the
derivation of the Feynman-Kac representation of the partition functions and of the off-diagonal
long-range order parameter. A complete account can be found in the excellent notes of
Ginibre \cite{Gin}; Faris wrote a useful survey \cite{Far}.

Let $D$ be the $d$-dimensional cubic box of size $L$ and volume $V=L^d$. We work with periodic boundary conditions, meaning that
$D$ is the $d$-dimensionial torus $\bbT_L^d$.
The state space is the Hilbert space $\caH_{D,N}$ of square-summable complex functions on $D^N$, that
are symmetric with respect to their arguments. Let $S$ denote the symmetric projector on
$L^2(D^N)$, i.e.
\be
S \psi(x_1,\dots,x_N) = \frac1{N!} \sum_{\pi\in S_N} \psi(x_{\pi(1)},\dots,x_{\pi(N)}),
\ee
where $x_1,\dots,x_N \in D$ and the sum is over all permutations of $N$ elements. The
state space for $N$ bosons in $D$ is therefore $\caH_{D,N} = S L^2(D^N)$, the projection
of $L^2(D^N)$ onto symmetric functions.

The Hamiltonian of the system is the sum $H=T+V$ of kinetic and interaction energies. The
kinetic energy is $T = -\sum_{j=1}^N \Delta_j$, where $\Delta_j$ is the Laplacian for the $j$-th particle. Interactions are given by the multiplication operator $V = \sum_{1\leq i<j\leq N} U(x_i-x_j)$.

Recall that $\beta$ and $\mu$ denote the inverse temperature and the chemical potential,
respectively. The canonical and grand-canonical partition functions are
\ba
&Y(\beta,V,N) = \Tr_{\caH_{D,N}} \e{-\beta H}, \\
&Z(\beta,V,\mu) = \sum_{N\geq0} \e{\beta\mu N} Y(\beta,V,N).
\label{gdcanfpart}
\end{align}
Under the assumption that $U(x)$ is a stable potential and that it
decays faster than $|x|^{-d}$ as $|x|\to\infty$, one can establish
the existence of the thermodynamic potentials (see \cite{Rue}) \ba
&\bsf(\beta,\rho) = \lim_{V\to\infty} -\frac1{\beta V} \log Y(N), \\
&\bsp(\beta,\mu) = \lim_{V\to\infty} \frac1V \log Z(\mu).
\end{align}
Further, $\bsf$ and $\bsp$ are related by a Legendre transform, \be
\label{Legendrerelation} \bsf(\beta,\rho) = \sup_\mu \bigl[ \rho\mu
- \tfrac1\beta \bsp(\beta,\mu) \bigr]. \ee This equation is useful
to find $\bsf$ from $\bsp$ in the case of the ideal gas, where
$\bsp$ can be computed explicitly.

The Feynman-Kac representation allows to express $\e{-\beta H}$ in terms of Wiener trajectories (Brownian motion). We briefly review the main properties of the Wiener measure.
Let $\caX_1$ be the set of continuous paths $\omega : [0,\beta] \to D$. Consider a function $F : \caX_1 \to
\bbR$ of the kind
\be
F(\omega) = f \bigl( \omega(t_1), \dots, \omega(t_n) \bigr),
\ee
where $f$ is a bounded measurable function on $D^n$, and $0<t_1<\dots<t_n<\beta$; we extend
$f$ on $\bbR^d$ by periodicity. The integral of $F$ with respect to the Wiener measure $W_{xy}^\beta$ is given by
\bm
\label{Wienergeneral}
\int_\caX F(\omega) \dd W_{xy}^\beta(\omega) = \sum_{z\in\bbZ^d} \int_{\bbR^{dn}} g_{t_1}(x_1-x)
g_{t_2-t_1}(x_2-x_1) \dots g_{\beta-t_n}(y+Lz-x_n) \\
f(x_1,\dots,x_n) \, \dd x_1 \dots \dd x_n,
\end{multline}
where $g_t$ is the normalized Gaussian function with mean zero and variance $2t$,
\be
g_t(x) = \frac1{(4\pi t)^{d/2}} \e{-\frac{x^2}{4t}}.
\ee
The sum over $z$ accounts for periodic boundary conditions. A special case of \eqref{Wienergeneral} is
when the function $F$ is the constant function $F(\omega) \equiv 1$; we get
\be
\label{Wienerperiodic}
\int\dd W^\beta_{xy}(\omega) = (4\pi\beta)^{-d/2} \sum_{z\in\bbZ^d} \e{-\frac{(x-y+Lz)^2}{4\beta}}.
\ee
Only the term $z=0$ remains in the limit $L\to\infty$.
It can be proved that such a measure exists and is unique \cite{Gin}. The Wiener measure $W_{xy}^{n\beta}$ is
concentrated on H\"older continuous trajectories (with any H\"older constant less than
$\frac12$) that start at $x$ and end at $y$. Integration with respect to $W_{xy}^\beta$ and $W_{00}^\beta$ are related as follows. Define $\omega'(t) = \omega(t) - t
\frac{y-x}\beta$; then
\be
\int F(\omega) \, \dd W_{xy}^\beta(\omega) = \e{-\frac{(y-x)^2}{4\beta}} \int F(\omega')
\, \dd W_{00}^\beta(\omega).
\ee

The Feynman-Kac formula states that $\e{-\beta H}$ is given by an integral operator
\cite{BR,Far,Gin}. We
are actually dealing with bosonic particles, and it is more convenient to consider the
operator $\e{-\beta H} S$ that also projects onto symmetric functions. We have
\be
\label{defintegraloperator}
\e{-\beta H} S \psi(x_1,\dots,x_N) = \int_{D^N} K(x_1,\dots,x_N;y_1,\dots,y_N)
\, \psi(y_1,\dots,y_N) \, \dd y_1 \dots \dd y_N,
\ee
where the kernel $K$ is given by
\begin{multline}
\label{defintegralkernel}
K(x_1,\dots,x_N;y_1,\dots,y_N) = \frac1{N!} \sum_{\pi\in S_N} \int \dd W^\beta_{x_1
y_{\pi(1)}}(\omega_1) \dots \dd W^\beta_{x_N y_{\pi(N)}}(\omega_N) \\
\exp\biggl\{ -\sum_{i<j} \int_0^\beta U \bigl( \omega_i(s) -
\omega_j(s) \bigr) \dd s \biggr\}.
\end{multline}
The canonical partition function is then given by
\bm
\label{fpartcanpi}
Y(N) = \sum_{\pi \in S_N} \frac1{N!} \int_{D^N} \dd x_1 \dots \dd x_N \int \dd
W^\beta_{x_1 x_{\pi(1)}}(\omega_1) \dots \dd W^\beta_{x_N x_{\pi(N)}}(\omega_N) \\
\exp \Bigl\{ -\sum_{i<j} \int_0^\beta U \bigl(
\omega_i(s)-\omega_j(s) \bigr) \dd s \Bigr\}.
\end{multline}
We now group the cycles into closed trajectories, that may wind several times around the
time direction. The number of permutations of $N$ elements with $k$ cycles of lengths $n_1,\dots,n_k$ (with $\sum_j n_j = N$) is
$$
\frac{N!}{k! \prod_j n_j}.
$$
Further, we have
\be
\label{concatenation}
\int_{D^{n-1}} \dd x_2 \dots \dd x_n \int\dd W_{x x_2}^\beta(\omega_1) \dots \dd W_{x_n y}^\beta(\omega_n) F(\omega) = \int\dd W_{xy}^{n\beta}(\omega) F(\omega).
\ee
The trajectory $\omega : [0,n\beta] \to D$ in the right side is the concatenation of $\omega_1,\dots,\omega_n$. The partition function \eqref{fpartcanpi} can then be rewritten in the form \eqref{FKrepcanpartfct}.

Let us turn to Penrose and Onsager off-diagonal long-range order \cite{PO}. Given a single
particle function $\varphi \in L^2(D)$, we define the operator $N_\varphi$ that
represents the number of particles in the state $\varphi$. The action of this operator is
given by
\be
\label{defNphi}
\bigl( N_\varphi \psi \bigr) (x_1,\dots,x_N) = \sum_{j=1}^N \int_D \overline{\varphi(x)}
\, \psi(x_1,\dots, \underbrace x_{j\text{-th place}}, \dots,x_N) \, \varphi(x_j) \, \dd x.
\ee
It is clear that $0 \leq N_\varphi \leq N$, and that $[N_\varphi,S]=0$. Let $\varphi_0(x)
\equiv \frac1{\sqrt V}$ denote the single particle ground state in absence of
interactions. It is also the Fourier function with mode $k=0$. The
average occupation of the zero mode is given by
\be
\label{defrho0}
\bsvarrho_\rho^{(0)} = \lim_{V\to\infty} \frac1{Y(N)} \Tr_{\caH_{D,N}}
\Bigl[ \frac{N_{\varphi_0}}V \e{-\beta H} \Bigr].
\ee
We set $N = \rho V$, and the limit exists at least along a subsequence of increasing volumes. A criterion for Bose-Einstein
condensation is that $\bsvarrho_\rho^{(0)}$ differs from zero. We can derive a Feynman-Kac
expression for this order parameter. From \eqref{defintegraloperator},
\eqref{defintegralkernel}, and \eqref{defNphi}, we have
\bm
\Tr_{\caH_{D,N}} N_\varphi \e{-\beta H} = \frac1{(N-1)!} \int_D \dd x
\, \overline{\varphi(x)} \int_D \dd y \, \varphi(y) \int_{D^{N-1}} \dd x_2 \dots \dd x_N
\sum_{\pi \in S_N} \\
\int \dd W^\beta_{x_1 \hat x_{\pi(1)}}(\omega_1) \dots \dd W^\beta_{x_N \hat
x_{\pi(N)}}(\omega_N) \exp \biggl\{ -\sum_{i<j} \int_0^\beta U \bigl( \omega_i(s) -
\omega_j(s) \bigr) \biggr\}.
\end{multline}
Here, we set $x_1=x$, $\hat x_1 = y$, and $\hat x_j = x_j$ for $2\leq j\leq N$. Then
$\bsvarrho_\rho^{(0)}$ can be written as
\be
\bsvarrho_\rho^{(0)} = \lim_{V\to\infty} \frac1{V^2} \int_{D^2} \sigma_\rho(x-y) \, \dd x
\, \dd y
\ee
where
\bm
\label{defodlro}
\sigma_\rho(x-y) = \frac1{Y(\beta,V,N)} \, \frac1{(N-1)!} \int_{D^{N-1}} \dd x_2 \dots
\dd x_N \sum_{\pi \in S_N} \\
\int \dd W^\beta_{x_1 \hat x_{\pi(1)}}(\omega_1) \dots \dd W^\beta_{x_N \hat
x_{\pi(N)}}(\omega_N) \exp \biggl\{ -\sum_{i<j} \int_0^\beta U \bigl( \omega_i(s) -
\omega_j(s) \bigr) \biggr\}.
\end{multline}
This expression involves an open cycle from $x$ to $y$, winding $n$ times around the time
direction, with $n=1,\dots,N$. Using the concatenation property \eqref{concatenation}, and
thanks to the combinatorial factor $\frac{(N-1)!}{(N-n)!}$, we obtain the expression
\eqref{odlro} for $\sigma_\rho(x-y)$.
The system displays off-diagonal long-range order if $\sigma_\rho(x)$ is strictly positive, uniformly in $V,x$.

\section{A simple proof of macroscopic occupation in the ideal gas}
\label{secgazparfait}

In this section, we give a proof of the macroscopic occupation of the zero mode at
low temperature. This is usually established in the grand-canonical ensemble, using a
chemical potential that varies with the volume and tends to zero in the thermodynamic
limit. This approach is rather unnatural, and requires large deviation techniques to control the fluctuations of the number of particles. The present proof is simpler and stays within the canonical ensemble.

The computation of the pressure and of the density in the
grand-canonical ensemble can be found in any textbook dealing with
quantum statistical mechanics. The chemical potential must be
strictly negative. The infinite volume pressure is \be
\bsp(\beta,\mu) = -\frac1{(2\pi)^d} \int_{\bbR^d} \log \bigl( 1 -
\e{-\beta (k^2 - \mu)} \bigr) \, \dd k, \ee and the density is \be
\label{rhoparfait} \bsrho(\beta,\mu) = \frac1{(2\pi)^d}
\int_{\bbR^d} \frac{\dd k}{\e{\beta (k^2 - \mu)} - 1} =
\frac1{(4\pi\beta)^{d/2}} \sum_{n\geq1} \e{\beta\mu n} n^{-d/2}. \ee
The limit of $\bsrho(\beta,\mu)$ as $\mu\nearrow0$ is finite for
$d\geq3$, and gives the {\it critical density} of the ideal Bose
gas, $\rho_{\rm c}$. The graph of $\bsp(\beta,\mu)$ in three
dimensions is plotted in Fig.\ \ref{figideal} (a). Its Legendre
transform \eqref{Legendrerelation} gives $\bsf(\beta,\rho)$, see
Fig.\ \ref{figideal} (b); it is nonanalytic at $\rho_{\rm c}$. The
value of $a(\beta)$ is given by \be a(\beta) = \lim_{\mu\nearrow0}
\tfrac1\beta \bsp(\beta,\mu) = \lim_{\rho\to\infty}
-\bsf(\beta,\rho). \ee

\bfig \epsfxsize=120mm \centerline{\epsffile{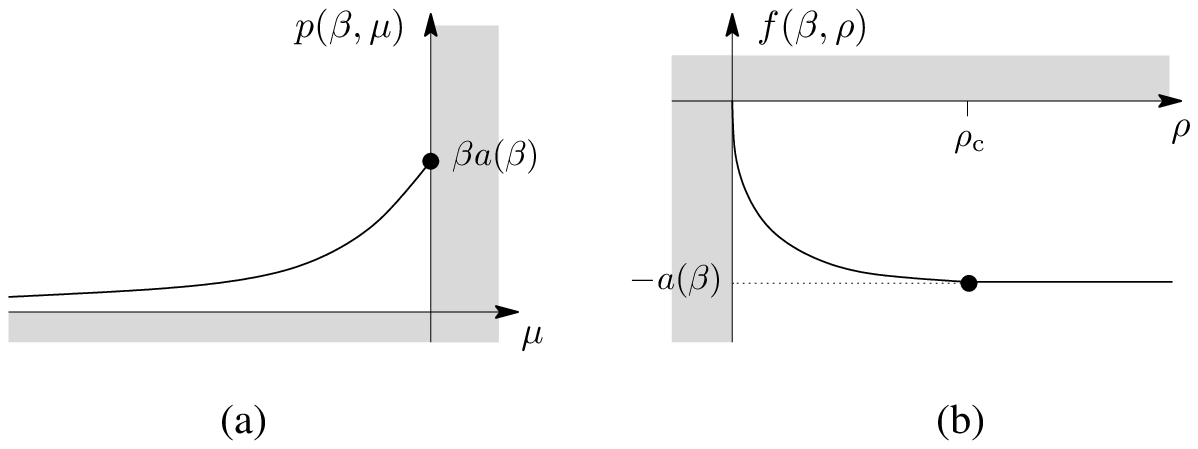}} \caption{The
pressure and the free energy of the ideal gas in three dimensions.}
\label{figideal}
\end{figure}

The very nature of the Bose-Einstein condensation is that the
occupation number for $k=0$ becomes macroscopic. The average
occupation of the zero mode $\bsvarrho^{(0)}_\rho$, see Eq.\
\eqref{defrho0}, can be rewritten as \be \bsvarrho^{(0)}_\rho =
\lim_{V\to\infty} \frac1{Y(N)} \sum_{(n_k): N} \frac{n_0}V \,
\e{-\beta \sum_k n_k k^2}. \ee Here, $N=V\rho$, and the sum is over
all occupation numbers $n_k \geq 0$, with indices $k \in
(\frac{2\pi}L \bbZ)^d$, such that $\sum_k n_k = N$. The heart of
Bose-Einstein condensation is the following result.

\begin{theorem}
\label{thmgazparfait}
For $d\geq3$, the single-particle ground state is macroscopically occupied if $\rho>\rho_{\rm c}$. More precisely,
$$
\bsvarrho^{(0)}_\rho = \max(0,\rho-\rho_{\rm c}).
$$
\end{theorem}

\begin{proof}
It is clear that $\bsvarrho^{(0)}_\rho \geq 0$.
We now establish that $\bsvarrho^{(0)}_\rho \geq \rho-\rho_{\rm c}$.
Let us introduce the average occupation of the mode $k$,
$$
\expval{n_k} = \frac1{Y(N)} \sum_{(n_{k'}):N} n_k \e{-\beta \sum_{k'} n_{k'}
{k'}^2}.
$$
Thanks to the sum rule $N =
\sum_k \expval{n_k}$, we have
\be
\label{rhozero}
\bsvarrho^{(0)}_\rho = \rho - \lim_{V\to\infty} \sum_{k\neq0} \frac{\expval{n_k}}V.
\ee
We can view $n_k$ as a random variable taking positive integer values; its expectation is
therefore given by
\be
\label{esperancenk}
\expval{n_k} = \sum_{i\geq1} {\rm Prob}(n_k \geq i),
\ee
where we defined
\be
{\rm Prob}(n_k \geq i) = \frac1{Y(N)} \sum_{(n_{k'}):N,i} \e{-\beta \sum_{k'}
n_{k'} {k'}^2}.
\ee
The sum is restricted to $(n_{k'})$ such that $\sum n_{k'} = N$ and $n_k \geq i$. The
change of variable $n_k \to n_k-i$ leads to
\be
\label{probank}
{\rm Prob}(n_k \geq i) = \e{-\beta i k^2} \frac{Y(N-i)}{Y(N)}.
\ee
The ratio of partition functions is also equal to the probability ${\rm Prob}(n_0 \geq
i)$, which is smaller than 1. Equations \eqref{esperancenk} and \eqref{probank} give a
bound for the occupation numbers of all modes $k\neq0$, namely,
\be
\expval{n_k} \leq \frac1{\e{\beta k^2} - 1}.
\ee
Notice that $\expval{n_k} \leq \frac1{\beta k^2} \leq \frac{L^2}{4\pi^2 \beta}$ for
$k\neq0$. This shows that only the zero mode can be macroscopically occupied (for
$d\geq3$). Inserting this bound into \eqref{rhozero}, we obtain
\be
\bsvarrho^{(0)}_\rho \geq \rho - \frac1{(2\pi)^d} \lim_{V\to\infty} \sum_{k\neq0} \Bigl( \frac{2\pi}L
\Bigr)^d \frac1{\e{\beta k^2} - 1}.
\ee
The limit converges to the expression \eqref{rhoparfait} with $\mu=0$, which is equal to
$\rho_{\rm c}$.

There remains to show that $\bsvarrho^{(0)}_\rho \leq
\max(0,\rho-\rho_{\rm c})$. From \eqref{probank} with $k=0$, and
using the equivalence of ensembles, we have for any fixed $a$, \be
\label{largedeviations} \lim_{V\to\infty} \frac 1{\beta V} \log {\rm
Prob}(n_0 \geq Va) = \bsf(\beta,\rho) - \bsf(\beta,\rho-a). \ee The
right side of \eqref{largedeviations} is strictly negative when $a >
\max(0,\rho-\rho_{\rm c})$. There exists $\delta>0$ such that for
large enough volumes, \be \label{uneborne} {\rm Prob}(n_0 \geq Va)
\leq \e{-V\delta}. \ee Let us assume that $\rho-\rho_{\rm c}>0$; the
case $\rho-\rho_{\rm c}\leq0$ can be treated similarly. Using
\eqref{esperancenk} with $k=0$, together with \eqref{uneborne}, we
get \be
\begin{split}
\frac{\expval{n_0}}V &= \frac1V \sum_{1\leq i\leq aV} {\rm
Prob}(n_0 \geq i) + \frac1V \sum_{aV <i\leq N} {\rm Prob}(n_0 \geq i) \\
&\leq a + \rho \e{-V\delta}.
\end{split}
\ee
It follows that $\bsvarrho^{(0)}_\rho$ is less than any number $a > \rho-\rho_{\rm c}$, hence
$\bsvarrho^{(0)}_\rho \leq \rho-\rho_{\rm c}$.
\end{proof}

\end{document}